\documentclass[preprint,showpacs,preprintnumbers,amsmath,amssymb]{revtex4}
\usepackage{graphicx}

\begin{document}
\def \ee {\varepsilon}
\thispagestyle{empty}
\title{Strengthening constraints on Yukawa-type
corrections to Newtonian gravity from measuring the Casimir force
between a cylinder and a plate}

\author{
G.~L.~Klimchitskaya\footnote{On leave from
{North-West Technical University, St.Petersburg,
 Russia}}
and C.~Romero
}

\affiliation{
Department of Physics, Federal University of Para\'{\i}ba,
C.P.5008, CEP 58059--900, Jo\~{a}o Pessoa, Pb-Brazil
}
\begin{abstract}
We discuss possibility to obtain stronger constraints on
non-Newtonian gravity from measuring the gradient of
the Casimir force between a cylinder and a plate.
 Exact analytical expression for the
Yukawa-type force in a cylinder-plate configuration is obtained,
as well as its asymptotic expansions. The gravitational force
is compared with the Casimir force acting between a cylinder
and a plate. Numerical computations for the prospective constraints
on non-Newtonian gravity are performed for recently proposed
experiment using a microfabricated cylinder attached to a
micromachined oscillator. Specifically, it is
shown that this experiment is expected to obtain up to
70 times stronger constraints on the Yukawa-type force,
compared with the best present day limits, over a wide
interaction range from 12.5 to 630\,nm.
\end{abstract}
\maketitle

\section{Introduction}

The Yukawa-type corrections to Newtonian gravitational law have
been discussed for a long time in connection with the problem of
the so-called {\it fifth force} \cite{1}. In the interaction of
two atoms belonging to different macrobodies such corrections
may arise due to the exchange of light hypothetical elementary
particles predicted in many extensions of the Standard Model
(scalar axion \cite{2}, graviphoton \cite{3}, dilaton \cite{4},
moduli \cite{5} etc.).  After the summation of atom-atom
Yukawa-type interactions over the volumes of two macrobodies,
one arrives at some force in addition to Newtonian gravity.

Interest in this kind of forces was rekindled after the
proposal of extra-dimensional models for which the compactification
energy can be as low as 1\,TeV \cite{6,7,8}. The respective
characteristic size of the compact manifold was shown to be
1\,mm and 5\,nm for two and three compact extra dimensions,
respectively \cite{8}. At separations several times larger than
the compactification scale it was predicted \cite{9,10} that
the standard Newton's gravitational law acquires the Yukawa-type
correction. Keeping in mind that at separations below $10\,\mu$m
corrections to Newton's law that are many orders of magnitude
larger than gravitation are not excluded experimentally, the
predicted correction has attracted considerable interest.
Specifically, a lot of gravitational experiments were performed
which imposed limits on the strength of Yukawa-type interaction
within short interaction ranges (see, for instance,
Refs.~\cite{11,12,13,14,15}). {}From these experiments,
strongest limits were obtained for Yukawa interaction ranges
larger than a few micrometers.

For interaction ranges of about one micrometer and less the
Casimir force occupies the place of Newtonian gravity as a
dominant background force. The Casimir force \cite{16}
originates from quantum fluctuations of the electromagnetic
field and finds diverse applications in both fundamental
physics and nanotechnology (see monographs \cite{17,18}).
Numerous experiments on measuring the Casimir force between
a sphere and a plate were  performed during the last few
years (see Ref.~\cite{19} for a review). The measure of
agreement between the experimental data and theory was used
to impose limits on the parameters of hypothetical
Yukawa-type interaction \cite{20,20a,20b,21,22,23,24,25,26}.
Eventually, the previously known constraints with
interaction ranges below one micrometer were strengthened
up to 10000 times.

Recently, the experiment on measuring the lateral Casimir
force between sinusoidally corrugated surfaces of a sphere
and a plate \cite{27,28} was used to constrain the
Yukawa-type force in the interaction range below 14\,nm.
Here, the measure of agreement between the data and the
exact theory based on the scattering approach with no
fitting parameters resulted \cite{29} in the strengthening
up to a factor of $2.4\times 10^{7}$. As opposed to
Ref.~\cite{21}, where the confidence level could not be
determined due to several uncertainties inherent to
relevant experiment \cite{30} on measuring the Casimir
force between two crossed cylinders, the constraints of
Ref.~\cite{29} were obtained at a 95\% confidence level.
In Ref.~\cite{29} a few prospective experiments were also
considered using the configurations of a plate and a
sphere and two parallel plates where surfaces of the test
bodies are smooth or covered with sinusoidal corrugations.
It was shown that some of these experiments are of high
promise for further strengthening of constraints on the
parameters of Yukawa-type hypothetical interaction.

Further prospects in the strengthening of constraints on
non-Newtonian gravity from the measurements of the Casimir
force are connected with the use of alternative configurations.
It was argued \cite{32a}, that the measurement setups using a
cylinder above a plate or two eccentric cylinders
combine some advantages of the
plate-plate and sphere-plate configurations.
Reference \cite{32a} noted that cylindrical geometries lead
to favorable conditions for the search of extra-dimensional
forces in the micrometer range.
In Ref.~\cite{31} the experimental setup with
a relatively large metallic cylinder of 6.35\,mm diameter
separated from the metallic plate with a gap of more than
$1\,\mu$m width was considered. Using the metal coated
cylindrical lens of even larger 24\,mm diameter,
serious difficulties have been found, however, in the
electrostatic calibration of Casimir apparatus \cite{32}.
Specifically, it was shown that the residual potential
difference between the grounded test bodies may become
dependent on the separation between them, and the force-distance
relation for the electric force deviates from the form
predicted by electrodynamics in a cylinder-plate
geometry. Similar anomalies were earlier found \cite{33}
in the sphere-plate geometry for spheres of centimeter-size
curvature radii (for spheres of about $100\,\mu$m radii anomalies
do not appear \cite{23,24,25,26,27,28,34}).
In Ref.~\cite{35} they were explained by unavoidable
deviations of mechanically polished and grounded
surfaces from perfect spherical shape assumed in
calculations. As recognized in Ref.~\cite{32}, deviations
of the surface from perfect cylindrical shape might be
also responsible for the calibration problems arising
in the case of a centimeter-size cylinder above a plate.
Because of this, in Ref.~\cite{36} the measurement of the
Casimir force between a microfabricated cylinder attached
to the micromechanical oscillator and a plate was
proposed. The respective setup uses cylinders with radii
of about $100\,\mu$m and gets the most benefit from the high
precision offered by a micromachined oscillator.

In this paper we investigate prospective constraints on
the parameters of Yukawa-type corrections to Newtonian
gravity which can be obtained from the measurements of
the Casimir force in the cylinder-plate geometry.
The Yukawa-type force between a cylinder and a plate is
calculated analytically both exactly and using the most
general form of the proximity force approximation (PFA),
i.e., the so-called {\it Derjaguin method} \cite{37},
with coinciding results. The respective expressions are
obtained both for homogeneous test bodies and for bodies
covered with layers made of different materials.
This allows immediate application of the obtained results
to the experiments in preparation.
The gravitational force acting between a cylinder and a plate
is calculated and compared with the Casimir force.
The range of parameters is determined where the gravitational
force can be discounted when obtaining constraints on the
Yukawa-type corrections to Newtonian gravity from the
measurements of the Casimir force between a cylinder and
a plate. The plate-based and cylinder-based versions of the
PFA \cite{38,39} are considered in application to the
calculation of the gravitational force.
Numerical computations are performed for the experiment using
a microfabricated cylinder attached to a micromachined
oscillator.
Prospective constraints on the parameters of Yukawa-type
corrections to Newtonian gravity that can be obtained
from the measurement of the Casimir force between a plate
and a microfabricated cylinder promise to be up to a factor
of 70 stronger than the previously known ones found in
Refs.~\cite{26,29}.
This provides further impetus to the planed experiments
on measuring the Casimir force in cylinder-plane geometry.
The obtained expressions for the Yukawa-type interaction
can be used not only for a microfabricated cylinder,
but in any other experiment exploiting the
cylinder-plate geometry.

The paper is organized as follows. In Sec.~II we calculate
the Yukawa-type force in the configuration of a cylinder above
a plate. Section~III is devoted to the calculation of the
gravitational force and to the comparison between
gravitational and Casimir forces. In Sec.~IV the prospective
constraints on the hypothetical Yukawa-type interaction are
derived. Section~V contains our conclusions and discussion.

\section{Yukawa-type force in cylinder-plate configuration}

We consider a circular cylinder of radius $R$, density $\rho_2$
and length $L$ situated above a large (i.e., with a dimension
much larger than $L$ and $2R$) plane plate of density $\rho_1$
and thickness $D_1$ parallel to it. Let the $z$-axis be perpendicular
to the plate, the upper surface of the plate be the plane
$z=0$ and the $y$-axis be parallel to the cylinder axis.
The separation distance between the plate and the cylinder is
denoted $a$.

The Yukawa interaction energy
between an atom of mass $m_2$ belonging to the
cylinder and an atom of mass $m_1$ at a separation $r$
belonging to the plate is conventially represented in the form
\begin{equation}
V_{\rm Yu}(r)=-\frac{Gm_1m_2}{r}\,\alpha\,
e^{-r/\lambda},
\label{eq1}
\end{equation}
\noindent
where $G$ is the Newtonian gravitational constant, $\alpha$ is
a dimensionless constant characterizing the strength of the Yukawa
interaction, and $\lambda$ is its interaction range.
Let us suppose that the atom $m_2$ is at a height $z$ above
the plate. Integration of Eq.~(\ref{eq1}) over the volume of
the plate and subsequent negative differentiation with
respect to $z$ results in the Yukawa force between the atom
$m_2$ and the plate \cite{40}
\begin{equation}
F_{\rm Yu}^{(m_2,p)}(z)=-2\pi Gm_2\rho_1\alpha\lambda\,
e^{-z/\lambda}\left(1-e^{-D_1/\lambda}\right).
\label{eq2}
\end{equation}
\noindent
To obtain the Yukawa-type force between a plate and a cylinder,
one should integrate Eq.~(\ref{eq2}) over the volume of the
cylinder. This leads to
\begin{eqnarray}
&&
F_{\rm Yu}^{(c,p)}(a)=-4\pi G\rho_1\rho_2\alpha\lambda L\,
\left(1-e^{-D_1/\lambda}\right)
\nonumber \\
&&~~~~~~~
\times\int_{a}^{a+2R} dz\,e^{-z/\lambda}
\left[R^2-(R+a-z)^2\right]^{1/2}
\nonumber \\
&&~~
=-4\pi G\rho_1\rho_2\alpha\lambda L\,
\left(1-e^{-D_1/\lambda}\right)e^{-a/\lambda}
\nonumber \\
&&~~~~~~~
\times\int_{0}^{2R} dv\,e^{-v/\lambda}
(2Rv-v^2)^{1/2},
\label{eq3}
\end{eqnarray}
\noindent
where the new integration variable $v=z-a$ was introduced.
Integrating by parts and using formula 3.366.1 in Ref.~\cite{41},
we arrive at the result
\begin{eqnarray}
&&
F_{\rm Yu}^{(c,p)}(a)=-4\pi^2 G\rho_1\rho_2\alpha\lambda^2 LR\,
\left(1-e^{-D_1/\lambda}\right)
\nonumber \\
&&~~~~~~~~~~~~
\times
e^{-(R+a)/\lambda}\,
{I}_1(R/\lambda),
\label{eq4}
\end{eqnarray}
\noindent
where ${I}_n(z)$ is the Bessel function of imaginary
argument.

The interaction ranges of the Yukawa-type forces discussed
below are $\lambda\leq 1\,\mu$m, whereas cylinder radii are
$R\geq 50\,\mu$m. This means that $R/\lambda\gg 1$ and
one can use the asymptotic expansion of large arguments
\cite{42}
\begin{equation}
{I}_1(R/\lambda)=\sqrt{\frac{\lambda}{2\pi R}}\,
e^{R/\lambda}\left(1+\tilde{u}_1\frac{\lambda}{R}+
\tilde{u}_2\frac{\lambda^2}{R^2}+\cdots\!\right),
\label{eq5}
\end{equation}
\noindent
where $\tilde{u}_1,\,\tilde{u}_2,\,\ldots$ are some numbers.
Substituting this in Eq.~(\ref{eq4}) and preserving only the
main contribution, we obtain the following expression for the
Yukawa-type force between a plate and a cylinder valid under
the condition $\lambda\ll R$:
\begin{equation}
F_{\rm Yu}^{(c,p)}(a)\approx
-2\pi G\rho_1\rho_2\alpha\lambda^2 L\,
\sqrt{2\pi R\lambda}\,e^{-a/\lambda}
\left(1-e^{-D_1/\lambda}\right).
\label{eq6}
\end{equation}
\noindent
It is interesting to compare the approximate result (\ref{eq6})
with the exact result (\ref{eq4}). Thus, at $\lambda/R$ equal
to 0.001, 0.003, 0.005, and 0.01, the relative difference between
the Yukawa-type forces computed using Eqs.~(\ref{eq4}) and
(\ref{eq6}) is equal to 0.038\%, 0.11\%, 0.19\%,  and 0.38\%,
respectively.

Taking into account that the Yukawa force is of potential nature,
its exact expression for any compact body above a plate
can be obtained \cite{40,43} by using the general formulation
of the PFA suggested by Derjaguin \cite{37}.
According to this formulation, the compact body is replaced
with a set of partial plane plates of appropriate thickness
$D_2(x)$ parallel to the underlying plate. Then the Yukawa
force can be calculated as
\begin{equation}
F_{\rm Yu,\,PFA}(a)=\int_{\sigma}P_{\,\rm Yu}(z)d\sigma,
\label{eq7}
\end{equation}
\noindent
where $P_{\rm Yu}(z)$ is the Yukawa pressure between two
infinite plane parallel plates of thicknesses $D_1$ and $D_2$
at a separation $z$, and $\sigma$ is the projection of a
compact body on the underlying plate (i.e., on the plane
$z=0$). Later, such an approach was called the {\it plate-based}
 PFA \cite{38,39}. The expression for $P_{\,\rm Yu}(z)$ can be
 simply obtained from Eq.~(\ref{eq2}) by integration over
 the volume of the upper plate of thickness $D_2$. It is
 given by \cite{40}
 \begin{eqnarray}
 &&
 P_{\,\rm Yu}(z)=-2\pi
 G\rho_1\rho_2\alpha\lambda^2\,e^{-z/\lambda}
 \nonumber \\
&&~~~~~~~~~~~~~~
\times
 \left(1-e^{-D_1/\lambda}\right)
\left(1-e^{-D_2/\lambda}\right),
\label{eq8}
\end{eqnarray}
\noindent
where for a cylinder above a plate the thickness
of a partial plate and its separation from the underlying plate
are given by
$
D_2(x)=2\sqrt{R^2-x^2}$,
$z(x)=a+R-\sqrt{R^2-x^2}$.
Substituting Eq.~(\ref{eq8}) into Eq.~(\ref{eq7})
we obtain
$
F_{\rm Yu,\,PFA}^{(c,p)}(a)=F_{\rm Yu}^{(c,p)}(a)$,
where $F_{\rm Yu}^{(c,p)}(a)$ is defined in Eq.~(\ref{eq4}),
i.e., the Yukawa force calculated by means of the plate-based
PFA coincides with the exact result, as it should be.

In experiments on measuring the Casimir force \cite{18,19}
the test bodies of densities $\rho_1$ and $\rho_2$ are
usually coated with two additional metallic layers.
Let the plate be coated with a layer of density
$\rho_1^{\prime}$ and thickness $\Delta_1^{\!\prime}$ and
with an outer layer of density $\rho_1^{\prime\prime}$ and
thickness $\Delta_1^{\!\prime\prime}$. In so doing $D_1$
is the plate thickness including layers. In a similar way,
we assume that the cylinder is coated with a layer of density
$\rho_2^{\prime}$ and thickness $\Delta_2^{\!\prime}$ and
with an outer layer of density $\rho_2^{\prime\prime}$ and
thickness $\Delta_2^{\!\prime\prime}$. The radius of the
cylinder coated with two layers is $R$.
Equation (4) for the Yukawa-type force can be applied to
a plate of thickness $\Delta_1^{\!\prime\prime}$ and
density $\rho_1^{\prime\prime}$ interacting with a cylinder
of radius $R-\Delta_2^{\!\prime}-\Delta_2^{\!\prime\prime}$
and density $\rho_2$ at a separation
$a+\Delta_2^{\!\prime}+\Delta_2^{\!\prime\prime}$ from the
plate. Then one should take into account the interaction of
the same plate $\Delta_1^{\!\prime\prime}$,
$\rho_1^{\prime\prime}$ with a cylindrical layer of
thickness $\Delta_2^{\!\prime}$ and density $\rho_2^{\prime}$.
This is achieved by applying Eq.~(\ref{eq4}) twice:
to a cylinder of radius $R-\Delta_2^{\!\prime\prime}$
and density $\rho_2^{\prime}$ at a separation
$a+\Delta_2^{\!\prime\prime}$ and to a cylinder of radius
$R-\Delta_2^{\!\prime}-\Delta_2^{\!\prime\prime}$ of the
same density at a separation
$a+\Delta_2^{\!\prime}+\Delta_2^{\!\prime\prime}$.
The obtained results should be subtracted. Using this
procedure, which is based on the additivity of Yukawa-type
force for a necessary number of times, one can simply determine the
contribution to the force of each covering layer.
Performing summation of all contributions, we arrive at the
exact Yukawa-type force acting between a plate and a cylinder
coated with two layers each:
\begin{equation}
F_{{\rm Yu},l}^{(c,p)}(a)=-4\pi^2G\alpha\lambda^2L\,e^{-a/\lambda}
\Phi_p\Phi_c.
\label{eq12}
\end{equation}
\noindent
Here, the following notations are introduced:
\begin{eqnarray}
&&
\Phi_p=\left[\rho_1^{\prime\prime}-(\rho_1^{\prime\prime}-
\rho_1^{\prime})\,e^{-\Delta_1^{\!\prime\prime}/\lambda}
\right.
\nonumber \\
&&~~~~~~~~
\left.
-(\rho_1^{\prime}-\rho_1)\,
e^{-(\Delta_1^{\!\prime}+\Delta_1^{\!\prime\prime})/\lambda}
-\rho_1\,e^{-D_1/\lambda}
\right],
\nonumber \\
&&
\Phi_c=e^{-R/\lambda}\,\left[
\vphantom{\left(\frac{\Delta_1^{\!\prime\prime}}{\lambda}\right)}
R\rho_2^{\prime\prime}\,
{I_1}\left(\frac{R}{\lambda}\right)\right.
\nonumber \\
&&~~~~~
-(\rho_2^{\prime\prime}-\rho_2^{\prime})
(R-\Delta_2^{\!\prime\prime})\,{I_1}
\left(\frac{R-\Delta_2^{\!\prime\prime}}{\lambda}\right)
\label{eq13} \\
&&~~~~
\left.
-(\rho_2^{\prime}-\rho_2)(R-\Delta_2^{\!\prime}-\Delta_2^{\!\prime\prime})
\,{I_1}\left(
\frac{R-\Delta_2^{\!\prime}-\Delta_2^{\!\prime\prime}}{\lambda}\right)
\right].
\nonumber
\end{eqnarray}

As was noted above, the interaction range $\lambda$ under
consideration is much less than the cylinder radius,
$\lambda\ll R$. Taking into account that the covering
layers are sufficiently thin, i.e., the inequalities
$R-\Delta_2^{\!\prime\prime}\gg\lambda$ and
$R-\Delta_2^{\!\prime}-\Delta_2^{\!\prime\prime}\gg\lambda$
are satisfied, the expression for $\Phi_c$ in Eq.~(\ref{eq13})
can be simplified
\begin{eqnarray}
&&
\Phi_c\approx\sqrt{\frac{\lambda}{2\pi}}\left[
\sqrt{R}\rho_2^{\prime\prime}
-\sqrt{R-\Delta_2^{\!\prime\prime}}
(\rho_2^{\prime\prime}-\rho_2^{\prime})
\,e^{-\Delta_2^{\!\prime\prime}/\lambda}\right.
\nonumber \\
&&~~~~
\left.
-\sqrt{R-\Delta_2^{\!\prime}-\Delta_2^{\!\prime\prime}}
(\rho_2^{\prime}-\rho_2)
\,e^{-(\Delta_2^{\!\prime}+\Delta_2^{\!\prime\prime})/\lambda}
\right].
\label{eq14}
\end{eqnarray}
\noindent
If, in addition, the conditions $\Delta_2^{\!\prime\prime}\ll R$
and $\Delta_2^{\!\prime}+\Delta_2^{\!\prime\prime}\ll R$
hold (this is, for instance, the case for experiments
on measuring the Casimir force in sphere-plate geometry), then
further simplification of Eq.~(\ref{eq14}) leads to
\begin{eqnarray}
&&
\Phi_c\approx\sqrt{\frac{\lambda R}{2\pi}}\left[
\rho_2^{\prime\prime}
-
(\rho_2^{\prime\prime}-\rho_2^{\prime})
\,e^{-\Delta_2^{\!\prime\prime}/\lambda}\right.
\label{eq15} \\
&&~~~~~~~~~~~~~\left.
-
(\rho_2^{\prime}-\rho_2)
\,e^{-(\Delta_2^{\!\prime}+\Delta_2^{\!\prime\prime})/\lambda}
\right].
\nonumber
\end{eqnarray}
\noindent
Equations (\ref{eq12})--(\ref{eq15}) allow computation of the
Yukawa-type correction to Newtonian gravity in the experimental
configuration of a cylinder above a plate. They will be used
in Sec.IV for the estimation of prospective constraints on
$\alpha$ and $\lambda$ which can be obtained from the
measurement of the Casimir force between a plate and a
microfabricated cylinder.

\section{Comparison of gravitational and Casimir forces}

As discussed in Sec.~I, strongest constraints on the Yukawa-type
interaction follow from measurements of the Casimir force at
interaction ranges where this force is the dominant background
force. Here, we compare the gravitational and Casimir forces
in the configuration of a cylinder and a plate and find where
the gravitational force is negligibly small.

The Newtonian potential of the gravitational force between atoms
$m_1$ and $m_2$ belonging to the plate and the cylinder,
respectively, is given by
\begin{equation}
V_{\rm G}=-\frac{Gm_1m_2}{r}
\label{eq16}
\end{equation}
\noindent
As in Sec.~II, we consider an atom $m_2$ at a height $z$ above
the plate. The gravitational force between an atom $m_2$ and
a plate is obtained \cite{40} by the integration of
Eq.~(\ref{eq16}) over the volume of the plate and subsequent
negative differentiation with respect to $z$
\begin{equation}
F_{\rm G}^{(m_2,p)}(z)=-2\pi G\rho_1m_2D_1.
\label{eq17}
\end{equation}
\noindent
It is seen that the gravitational force does not depend on $z$.
The gravitational force between a plate and a cylinder is
obtained from the integration of Eq.~(\ref{eq17}) over the
cylinder volume
\begin{equation}
F_{\rm G}^{(c,p)}=-2\pi^2 G\rho_1\rho_2D_1LR^2.
\label{eq18}
\end{equation}

The same result is obtainable with the help of the plate-based
PFA. Considering the second plate of thickness $D_2$ and
density $\rho_2$ parallel to the first, the gravitational
pressure between the two plates can  be obtained \cite{40}
using Eq.~(\ref{eq17})
\begin{equation}
P_{\,\rm G}=-2\pi G\rho_1\rho_2D_1D_2.
\label{eq19}
\end{equation}
\noindent
Then the PFA result for the gravitational force between a plate
and a cylinder is found from Eq.~(\ref{eq7}), where $P_{\,\rm Yu}$
should be replaced with $P_{\,\rm G}$
\begin{eqnarray}
F_{\rm G,\,PFA}^{(c,p)}&=&-4\pi G\rho_1\rho_2D_1L
\int_{0}^{R}dxD_2(x)
\nonumber \\
&=&
-2\pi^2 G\rho_1\rho_2D_1LR^2.
\label{eq20}
\end{eqnarray}
\noindent
Here, $D_2(x)$ was used as defined below Eq.~(\ref{eq8}). {}From
Eq.~(\ref{eq20}) it can be seen that the PFA result obtained
using the Derjaguin method coincides with the exact
result (\ref{eq18}).
Note that the cylinder-based PFA does not lead to correct
results for the gravitational force [specifically, for a cylinder
above a plate a factor of two larger force magnitude is obtained
using this method, as compared with Eq.~(\ref{eq20})].

The Casimir force acting between a metal coated
cylinder and a metal coated plate in thermal equilibrium at
temperature $T$ was calculated in Ref.~\cite{36} using the
plate-based PFA. It is given by
\begin{eqnarray}
&&
F_{\rm C}^{(c,p)}(a,T)=-\frac{k_BTL}{4\sqrt{\pi} a^2}\sqrt{\frac{R}{2a}}
\sum\limits_{l=0}^{\infty}{\vphantom{\sum}}^{\prime}
\int_{\tau l}^{\infty}v^{3/2}\,dv
\nonumber \\
&&~~~~\times
\left[{\rm Li}_{1/2}(r_{\rm TM}^{2}e^{-v})+
{\rm Li}_{1/2}(r_{\rm TE}^{2}e^{-v})\right].
\label{eq22}
\end{eqnarray}
\noindent
Here, $k_B$ is the Boltzmann constant, ${\rm Li}_n(z)$ is the
polylogarithm function, $\tau=4\pi k_BTa/(\hbar c)$,
$l=0,\,1,\,2,\,\ldots\,$, and prime near the summation sign adds
a multiple 1/2 to the term with $l=0$. The reflection coefficients
for the two independent polarizations of the electromagnetic field
(transverse magnetic, TM, and transverse electric, TE) are
given by
\begin{eqnarray}
&&
r_{\rm TM}=
\frac{\varepsilon_lv-\sqrt{v^2+(\varepsilon_l-1)\tau^2l^2}}{\varepsilon_lv+
\sqrt{v^2+(\varepsilon_l-1)\tau^2l^2}},
\nonumber \\
&&
r_{\rm TE}=
\frac{v-\sqrt{v^2+(\varepsilon_l-1)\tau^2l^2}}{v+
\sqrt{v^2+(\varepsilon_l-1)\tau^2l^2}},
\label{eq23}
\end{eqnarray}
\noindent
where the dielectric permittivity
$\varepsilon_l=\varepsilon(i\xi_l)$ is calculated at the imaginary
Matsubara frequencies $\xi_l=\tau lc/(2a)$.
For Au coated bodies used in experiments on measuring the
Casimir force $\varepsilon(\omega)$ was described by means
of the experimentally consistent generalized plasma-like
model \cite{18,19,26,36} with the plasma frequency
$\omega_p=9.0\,$eV \cite{45} (note that if a plate and a
cylinder are coated with metallic layers, only external
layers of thicknesses $\Delta_1^{\!\prime\prime}$,
$\Delta_2^{\!\prime\prime}$ determine the value of the
Casimir force). The use of some alternative model of
dielectric permittivity of Au (for instance, on the basis
of tabulated optical data extrapolated to low frequencies
by means of the Drude model) leads to almost the same result
with respect to the comparison between Casimir and gravitational
forces at separations below a micrometer.

Now we present a few computational results for the gravitational
force and for the ratio of gravitational to Casimir forces.
First, we consider the plate of $D_1=5\,$mm thickness and
cylinders of different radii made of bulk Au with the density
$19.28\times 10^{3}\,\mbox{kg/m}^3$.
In this case, the values of the gravitational force per unit
cylinder length calculated using Eq.~(\ref{eq18}) are equal
to 6.12, 24.47, and 55.06\,pN/m for cylinder radii 50, 100, and
$150\,\mu$m, respectively. The ratio
$F_{\rm G}^{(c,p)}/F_{\rm C}^{(c,p)}$ is equal to
$5.24\times 10^{-4}$, $1.48\times 10^{-3}$, and
$2.72\times 10^{-3}$ with the same respective values of
cylinder radii at the plate-cylinder separation distance
$a=1\,\mu$m. This ratio increases with the increase of $a$.
For instance, at $a=2\,\mu$m it is equal to
$5.29\times 10^{-3}$, $1.50\times 10^{-2}$, and
$2.75\times 10^{-2}$ for cylinder radii 50, 100, and
$150\,\mu$m, respectively.
In Fig.~1(a) we present the values of the ratio
$F_{\rm G}^{(c,p)}/F_{\rm C}^{(c,p)}$  as a function of
separation for cylinder radii $R=50\,\mu$m (line 1),
$100\,\mu$m (line 2), and $150\,\mu$m (line 3). It is still
assumed that both the cylinder and the plate are made
of bulk Au. As can be seen in Fig.~1(a), for microfabricated
cylinders the gravitational force achieves some noticeable
fraction of the Casimir force only at separations of a few
micrometers.

In real experiments involving the layer structure of the test
bodies the role of the gravitational force, as compared to the
Casimir force, is even less. To illustrate this, we consider
the generalization of Eq.~(\ref{eq18}) for the case of test
bodies coated with two metal layers each (see Sec.~II for
all notations). Using the additivity of the gravitational
force, the role of layers can be accounted for in the same
way as it was done for the Yukawa-type force in Sec.~II.
As a result, Eq.~(\ref{eq18}) is replaced with
\begin{eqnarray}
&&
F_{{\rm G},\,l}^{(c,p)}=-2\pi^2 GL
\left[\rho_1^{\prime\prime}\Delta_1^{\!\prime\prime}+
\rho_1^{\prime}\Delta_1^{\!\prime}+\rho_1(D_1-
\Delta_1^{\!\prime}-\Delta_1^{\!\prime\prime})\right]
\nonumber \\
&&~~~~~~\times\left[
\rho_2^{\prime\prime}R^2-(\rho_2^{\prime\prime}-\rho_2^{\prime})
(R-\Delta_2^{\!\prime\prime})^2\right.
\label{eq24} \\
&&~~~~~~~~~~~~~\left.
-(\rho_2^{\prime}-\rho_2)
(R-\Delta_2^{\!\prime}-\Delta_2^{\!\prime\prime})^2\right].
\nonumber
\end{eqnarray}
\noindent
For numerical calculations we consider the same layer structure
of the test bodies as in Refs.~\cite{25,26}, i.e., Si plate coated
with layers of Cr and Au and sapphire cylinder also coated with
Cr and Au layers. The respective densities and thicknesses are
the following:
$\rho_1^{\prime\prime}=\rho_2^{\prime\prime}=19.28\times 10^{3}
\,\mbox{kg/m}^3$,
$\rho_1^{\prime}=\rho_2^{\prime}=7.14\times 10^{3}\,\mbox{kg/m}^3$,
$\rho_1=2.33\times 10^{3}\,\mbox{kg/m}^3$,
$\rho_2=4.1\times 10^{3}\,\mbox{kg/m}^3$,
$\Delta_1^{\!\prime\prime}=210\,$nm, $\Delta_2^{\!\prime\prime}=180\,$nm,
$\Delta_1^{\!\prime}=\Delta_2^{\!\prime}=10\,$nm, and $D_1=5\,$mm.
The computational results for the ratio
$F_{{\rm G},\,l}^{(c,p)}/F_{\rm C}^{(c,p)}$
in the case of test bodies with layer structure are shown in
Fig.~1(b) as a function of separation over the region from 1 to
$5\,\mu$m (below $1\,\mu$m this ratio is vanishingly small).
Lines 1, 2, and 3 are plotted for cylinder radii $R=50$, 100
and $150\,\mu$m, respectively. As can be seen in Fig.~1(b),
for a cylinder and a plate with layer structure used in real
experiments on measuring the Casimir force the ratio of
gravitational and Casimir force is much less than for bulk Au
test bodies. Even at a separation $a=5\,\mu$m and for largest
cylinder radius $R=150\,\mu$m this ratio is less than 0.01.
This allows us to conclude that in the interaction range below
$1\,\mu$m under consideration in the next section the
gravitational force between a plate and a microfabricated
cylinder does not play any role and the Yukawa-type correction
to gravity should be considered in the background of the
Casimir force.

\section{Prospective constraints}

In this section we estimate the strength of constraints on the
parameters of Yukawa-type corrections to Newtonian gravity
that can be obtained from the measurements of the Casimir force
between a microfabricated cylinder and a plate proposed in
Ref.~\cite{36}. Similar to already performed experiments of
Refs.~\cite{23,24,25,26}, the experiment of Ref.~\cite{36} is
dynamic, which is to say that the separation distance between the
cylinder and the plate is varied harmonically at the natural
frequency of the micromachined oscillator. The immediately
measured quantity is the shift of this frequency under the
influence of the Casimir force. In its turn, from the
solution of a simple mechanical problem it follows that this
shift of the resonance frequency is proportional to the
gradient of the Casimir force \cite{46,47}.

Below we assume that the confidence interval for the difference
between theoretical and experimental gradients of the Casimir
force in cylinder-plate configuration determined at a 95\%
confidence level $[-\Theta(a),\Theta(a)]$ is the same
as in Ref.~\cite{26}. Keeping in mind that the half width
of the confidence interval $\tilde{\Xi}(a)$ in Ref.~\cite{26}
was determined for the difference of the Casimir pressure
between two parallel plates, which is connected with the
gradient of the Casimir force in sphere-plate geometry
by the equation
\begin{equation}
P_{\,\rm C}(a)=-\frac{1}{2\pi R}\,
\frac{\partial F_{\rm C}^{(s,p)}(a)}{\partial a},
\label{eq25}
\end{equation}
\noindent
we arrive at the equality $\Theta(a)=2\pi R\tilde{\Xi}(a)$,
where $\tilde{\Xi}(a)$ at different separations is specified
in Ref.~\cite{26}. A cylinder of $R=151.3\,\mu$m radius
is used, i.e., one of the same radius as
that of the sphere in Ref.~\cite{26}.
The length of the cylinder $L=\pi R/2$ is found from the
condition that its projection on the plate is the same as the
projection of the sphere.

Prospective constraints on the Yukawa-type corrections to
Newtonian gravity can be estimated from the assumption that
in the limit of the confidence interval $\Theta(a)$ the
hypothetical Yukawa interaction will not be observed.
This leads to the inequality
\begin{equation}
\left|
\frac{\partial F_{{\rm Yu},\,l}^{(c,p)}(a)}{\partial a}
\right|\leq\Theta(a),
\label{eq26}
\end{equation}
\noindent
where the gradient of the Yukawa force can be found from
Eq.~(\ref{eq12}):
\begin{equation}
\frac{\partial F_{{\rm Yu},\,l}^{(c,p)}(a)}{\partial a}
=4\pi^2G\alpha\lambda L\,e^{-a/\lambda}\,\Phi_p\Phi_c.
\label{eq27}
\end{equation}
\noindent
The exact expressions for the factors $\Phi_p$ and $\Phi_c$
are presented in Eq.~(\ref{eq13}), and the asymptotic
representations for $\Phi_c$ in the case of large cylinder
radii, thin layers and short interaction ranges in
Eqs.~(\ref{eq14}) and (\ref{eq15}).

The prospective constraints on $\lambda$ and $\alpha$
following from Eqs.~(\ref{eq26}) and (\ref{eq27}) are
shown in Fig.~2 by the dashed line, where the region of
$(\lambda,\,\alpha)$-plane above the line  is expected to
be prohibited from the planed measurements of the gradient
of the Casimir force between a cylinder and a plate and
the region below this line is allowed.
Note that strongest constraints shown in Fig.~2 are obtained at
different separations $a$. Thus, in the region
$10\,\mbox{nm}\leq\lambda\leq 40\,$nm the constraints shown by the dashed
line are obtained at $a=180\,$nm, in the region
$50\,\mbox{nm}\leq\lambda\leq 63\,$nm at $a=200\,$nm etc.
For $\lambda\leq 315\,$nm the asymptotic expressions (\ref{eq14})
and (\ref{eq15}) for the factor $\Phi_c$ lead to the same
constraints as exact Eq.~(\ref{eq13}).
As noted in Ref.~\cite{44}, the half width of the confidence
interval is determined not more precisely than with two or,
at maximum, three significant figures. Because of this,
independently of how precise the measurements of the Casimir
force are, the Yukawa force never needs to be calculated
with a higher than up to 0.1\% accuracy.

For comparison purposes, in Fig.~2 we also plot the strongest
constraints on the parameters of Yukawa-type interaction obtained
up to now (solid lines 1--4). Line 1 is obtained in
Ref.~\cite{29} from measurements of the lateral Casimir force
between corrugated surfaces \cite{27,28}. Line 2 follows from
dynamic determination of the Casimir pressure by means of a
micromechanical oscillator \cite{26,44}. Line 3 is plotted
using the results of Casimir-less experiment \cite{48} and
line 4 is obtained \cite{20} from the results of torsion pendulum
experiment \cite{49}. As can be seen in Fig.~2, the strongest
strengthening of presently known constraints up to 70 times
can be achieved at $\lambda=18\,$nm. At $\lambda=80$ and 422\,nm
the expected strengthening of constraints is by a factor 28 and
7, respectively. All of all, the proposed experiment promises
to strengthen the presently known constraints over the very wide
interaction range $12.5\,\mbox{nm}\leq\lambda\leq 630\,$nm.

\section{Conclusions and discussion}

In the foregoing we have investigated the potentialities of
experiments on measuring the Casimir force in cylinder-plate
geometry for strengthening constraints on Yukawa-type corrections
to Newtonian gravity. The configuration of a cylinder above a
plate has long attracted attention with respect to
measurements of the Casimir force \cite{32a,31,32,51a,51b}. Interest to it
has  rekindled after the suggestion \cite{36} to use
microfabricated cylinders and the setup of a micromachined
oscillator which allowed to achieve highest experimental precision
for the configuration of a sphere and a plate.

Here, we have obtained exact analytical expression for the
Yukawa-type force acting between a cylinder and
a large plate. This expression is generalized for the case of
test bodies coated with thin metallic layers, as is the case
in typical experiments on measuring the Casimir force and its
gradient. It was shown that the same expression can be obtained
using the Derjaguin formulation of the PFA.
The magnitude of the gravitational force acting between a cylinder
and a plate was compared with the magnitude of the Casimir force,
and the former was shown to be negligibly small at the separations
under consideration. With respect to the gravitational force,
the plate-based and the cylinder-based versions of the PFA were
discussed and found not to be equivalent. Starting from the
assumption that main parameters of the proposed experiment on
the measurement of the gradient of the Casimir force between a
cylinder and a plate are the same as in the experiment of
Ref.~\cite{26}, we have estimated the prospective constraints on
the Yukawa-type interaction obtainable using a microfabricated
cylinder attached to a micromachined oscillator.
It was shown that the strengthening of constraints up
to a factor 70 is expected over the wide interaction range from
$\lambda=12.5\,$nm to $\lambda=630\,$nm.
The obtained expressions for the Yukawa interaction can be used
in all experiments exploiting the cylinder-plate geometry.
At the moment, the
strongest constraints valid within this interaction region are
obtained from four different experiments on measuring the
Casimir force.

The presented above results show that the Casimir effect
has potentialities as
the source of stronger constraints on non-Newtonian
gravity. While the cylinder-plate configuration considered
here promises obtaining stronger constraints in the
interaction range below $1\,\mu$m, it might be expected that
the proposed experiment with two parallel plates coated with
gold \cite{50} will result in stronger constraints on
the Yukawa force in the interaction range of a few micrometers.

\section*{Acknowledgments}

G.L.K.\ is grateful to the Federal University of
Para\'{\i}ba (Jo\~{a}o Pessoa, Brazil), where this work was
performed, for kind hospitality.  The authors
acknowledge CNPq for financial
support.


\begin{figure}[b]
\vspace*{-5cm}
\centerline{\hspace*{1cm}
\includegraphics{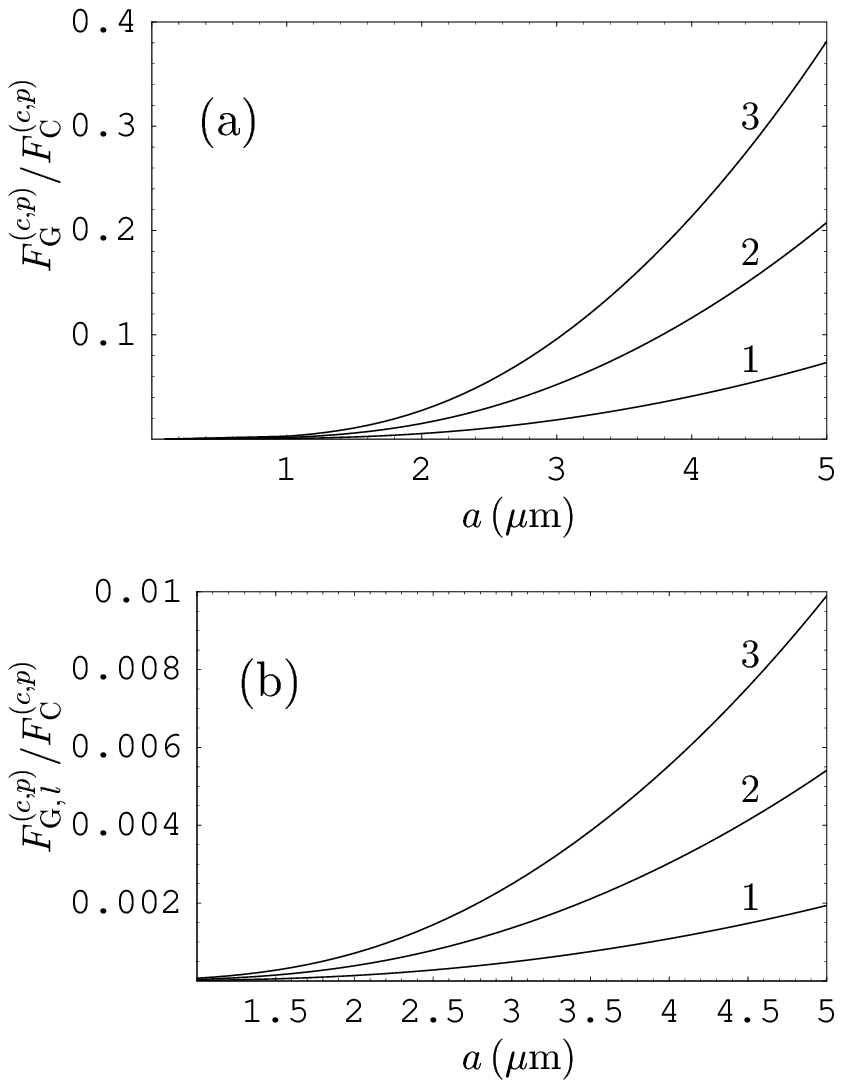}
}
\vspace*{-11cm}
\caption{
Ratio between the gravitational and Casimir forces in
a cylinder-plate configuration (a) for bulk gold bodies
and (b) for bodies coated with thin metallic layers
(see text for further details). The thickness of the
plate is 5\,mm. The computational results for cylinder
radii $R=50$, 100, and $150\,\mu$m are shown by the
lines 1, 2, and 3, respectively.
}
\end{figure}
\begin{figure}[b]
\vspace*{-11cm}
\centerline{\hspace*{3cm}
\includegraphics{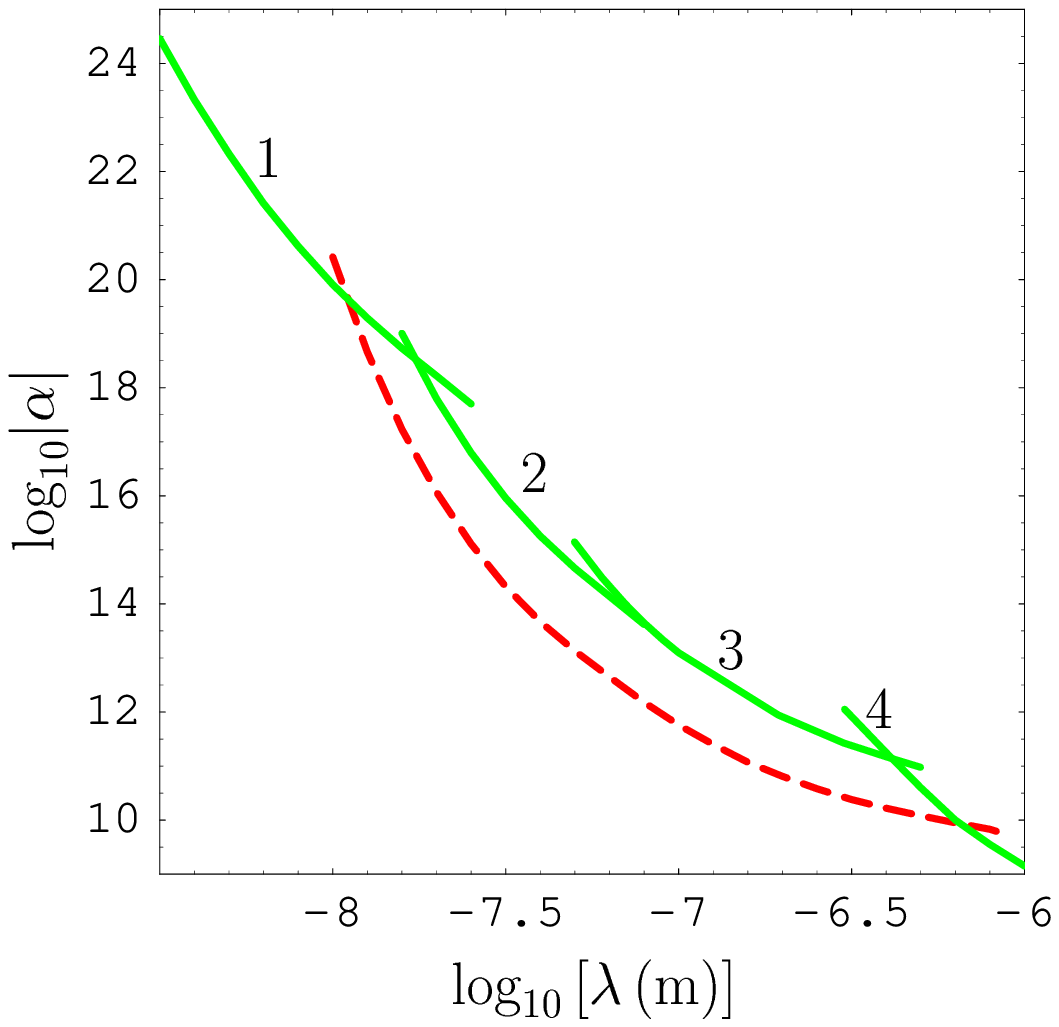}
}
\vspace*{-6cm}
\caption{(Color online)
The strongest currently available
constraints on the parameters of Yukawa-type force
are shown by the solid lines 1--4 (see text for further
discussion).
The dashed line shows the prospective constraints that
can be obtained from measuring the gradient of the
Casimir force between a plate and a microfabricated
cylinder.
The allowed regions in the $(\lambda,\,\alpha)$-plane lie
beneath the lines.
}
\end{figure}
\end{document}